\documentclass[11pt,a4paper]{article}
\usepackage{jcappub}
\usepackage{graphicx}
\usepackage{amssymb}
\usepackage{amsmath}
\usepackage{bm}
\usepackage{latexsym}
\usepackage{epsfig}
\usepackage{psfrag}
\usepackage[normalem]{ulem}
\usepackage{textcomp}
\usepackage{color}
\usepackage{pstricks}
\usepackage[utf8]{inputenc}
\makeatletter
\gdef\@fpheader{}
\makeatother

\def\nn{\nonumber} 
\def\pa{{\partial}}
\def\f{\frac}
\def\l{\left}
\def\r{\right}
\def\d{{\rm d}}
\def\Mpl{M_{_{\rm Pl}}}
\def\beq{\begin{equation}}
\def\eeq{\end{equation}} 
\def\beqa{\begin{eqnarray}}
\def\eeqa{\end{eqnarray}}

\def\vk{{\bm k}}

\def\vka{{\bm k}_{1}}
\def\vkb{{\bm k}_{2}}
\def\vkc{{\bm k}_{3}}
\def\ska{{k_{1}}}
\def\skb{{k_{2}}}
\def\skc{{k_{3}}}
\def\Mp{M_{_{\rm Pl}}}
\def\cG{{\cal G}}
\def\ei{\eta_{\rm i}}

\def\ee{\eta_{\rm e}}
\def\hnl{h_{_{\rm NL}}}

\def\nt{n_{_{\rm T}}}
\def\pt{{\mathcal P}_{_{\rm T}}}

\def\cB{{\mathcal B}}

\newcommand{\g}{\gamma}

\newcommand{\viz}{\textit{viz.~}}
\newcommand{\ie}{\textit{i.e.~}}

\begin{document}
%%%%%%%%%%%%%%%%%%%%%%%%%%%%%%%%%%%%%%%%%%%%%%%%%%%%%%%%%%%%%%%%%%%%%%%%%%%%%%%
\title{Numerical evaluation of the tensor bispectrum in two field inflation}
\author{Rathul Nath Raveendran}
\affiliation[a]{The Institute of Mathematical Sciences, HBNI, CIT Campus,
Chennai~600113, India}
\emailAdd{rathulnr@imsc.res.in}
\author{and L.~Sriramkumar}
\affiliation[b]{Department of Physics, Indian Institute of 
Technology Madras, Chennai~600036, India}
\emailAdd{sriram@physics.iitm.ac.in}
\date{today} 
%%%%%%%%%%%%%%%%%%%%%%%%%%%%%%%%%%%%%%%%%%%%%%%%%%%%%%%%%%%%%%%%%%%%%%%%%%%%%%%
\abstract{We evaluate the dimensionless non-Gaussianity parameter $\hnl$, 
that characterizes the amplitude of the tensor bispectrum, 
{\it numerically}\/ for a class of two field inflationary models such as 
double inflation, hybrid inflation and aligned natural inflation. 
We compare the numerical results with the slow roll results which can be 
obtained analytically. 
In the context of double inflation, we also investigate the effects on $\hnl$
due to curved trajectories in the field space. 
We explicitly examine the validity of the consistency relation governing the 
tensor bispectrum in the squeezed limit.
Lastly, we discuss the contribution to $\hnl$ due to the epoch of preheating 
in two field models.}
\maketitle
%%%%%%%%%%%%%%%%%%%%%%%%%%%%%%%%%%%%%%%%%%%%%%%%%%%%%%%%%%%%%%%%%%%%%%%%%%%%%%%

%%%%%%%%%%%%%%%%%%%%%%%%%%%%%%%%%%%%%%%%%%%%%%%%%%%%%%%%%%%%%%%%%%%%%%%%%%%%%%%

\section{Introduction}

In the absence of equally effective alternatives, the inflationary paradigm 
continues to remain the most compelling scenario to describe the origin of 
perturbations in the primordial universe.
Inflation---which refers to a period of accelerated expansion during the 
early stages of the radiation dominated epoch---was initially proposed to 
explain cosmological observations such as the extent of homogeneity and 
spatial flatness of the universe. 
However, soon after the original proposal, it was realized that apart from
helping to overcome the drawbacks of the conventional hot big bang model, the 
inflationary scenario also provides a causal mechanism for the generation
of primordial perturbations. 
According to the inflationary paradigm, the primordial perturbations are 
generated due to quantum fluctuations, which are rapidly stretched to 
cosmological scales due to the accelerated expansion. 
The perturbations generated during inflation lead to anisotropies in the 
Cosmic Microwave Background (CMB), which in turn result in the large 
scale structure of galaxies and clusters of galaxies that we see around 
us today (see, for instance, any of the following 
reviews: Refs.~\cite{Mukhanov:1990me,Martin:2003bt,Bassett:2005xm,
Sriramkumar:2009kg,Baumann:2009ds,Sriramkumar:2012mik,Martin:2015dha}).

\par

Typically, the period of accelerated expansion is assumed to be driven by 
scalar fields. 
Many models consisting of single and multiple scalar fields have been 
proposed to achieve inflation.
The potentials governing the scalar fields, along with the values of the 
parameters describing them, determine the dynamics during inflation.
It is the quantum fluctuations associated with the scalar fields that are
responsible for the primordial perturbations. 
The background inflationary dynamics determines the characteristics 
of these perturbations, which are conveniently described in terms of 
correlation functions.
The CMB and other cosmological data point to a nearly scale invariant 
primordial scalar power spectrum as is generated by the simplest models 
of slow roll inflation~\cite{Mortonson:2010er,Easther:2011yq,Norena:2012rs,
Martin:2010hh,Martin:2013tda,Martin:2013nzq,Martin:2014lra,Martin:2014rqa,
Ade:2015lrj}. 
However, despite the strong constraints that have emerged, there exist many 
inflationary models that are consistent with the data at the level of 
two-point functions.
In the case of canonical single field models, there has been a comprehensive 
comparative analysis of a fairly large set of models with the cosmological
data~\cite{Martin:2010hh,Martin:2013tda,Martin:2013nzq,Martin:2014lra,
Martin:2014rqa,Ade:2015lrj}.
Clearly, in the long run, it would be desirable to carry out a similar 
comparison of multi field models and, more specifically, two field models
with the data (see, for instance, 
Refs.~\cite{Tsujikawa:2002qx,Parkinson:2004yx,Easther:2013rva,Ade:2015lrj}).
As far as the background evolution is concerned, two field models offer 
a richer dynamics than the single field models due to the possibility of 
different types of trajectories in the field space.
At the level of perturbations, the existence of iso-curvature 
perturbations in multi field models can lead to a non-trivial evolution of the 
curvature perturbation on super-Hubble scales (see, for example, the following 
articles~\cite{Polarski:1994bk,Langlois:1999dw,Gordon:2000hv,Bartolo:2001rt,
Byrnes:2006fr,Lalak:2007vi,Langlois:2008mn} or 
reviews~\cite{Wands:2007bd,Langlois:2010xc,Gong:2016qmq}).

\par

Over the last decade and a half, it has been recognized that observations
of primordial non-Gaussianities---in particular, the amplitude of  
three-point functions---can help us arrive at a smaller class of viable 
inflationary models.
This expectation has been corroborated to a large extent by the strong 
constraints that have been arrived at by the Planck data on the three
non-Gaussianity parameters that describe the amplitude of the scalar 
bispectrum~\cite{Ade:2013ydc}.
Theoretically, a considerable amount of work that has been carried out 
towards understanding the non-Gaussianities generated in single and 
multi field inflationary models.
However, the theoretical understanding of non-Gaussianities generated in  
inflationary models and the observational constraints that have been arrived 
at are largely concentrated on the scalar bispectrum and the corresponding 
non-Gaussianity parameters~\cite{Maldacena:2002vr,Seery:2005wm,Chen:2005fe,
Chen:2006nt,Langlois:2008wt,Langlois:2008qf,Chen:2010xka,Wang:2013zva,Komatsu:2001rj,
Komatsu:2003iq,Babich:2004yc,Liguori:2005rj,Hikage:2006fe,Fergusson:2006pr,
Yadav:2007rk,Creminelli:2006gc,Yadav:2007yy,Hikage:2008gy,Rudjord:2009mh,
Smith:2009jr,Smidt:2010ra,Fergusson:2010dm}.

\par

In fact, apart from the scalar bispectrum, there arise three other three-point 
functions when the tensor perturbations are also included~\cite{Maldacena:2011nz,
Gao:2011vs,Gao:2012ib,Sreenath:2013xra}.
The three-point functions are often evaluated analytically in the slow roll 
approximation, and one has to resort to numerical efforts to evaluate these 
three-point functions in a generic situation (in this context, see, for instance, 
Refs.~\cite{Chen:2006xjb,Chen:2008wn,Adshead:2011bw,Adshead:2011jq,Hazra:2012yn,
Sreenath:2013xra,Sreenath:2014nka,Sreenath:2014nca}). 
Also, while numerical procedures have been developed to evaluate the three-point
functions in single field models~\cite{Chen:2006xjb,Chen:2008wn,Adshead:2011bw,
Adshead:2011jq,Hazra:2012yn,Sreenath:2013xra,Sreenath:2014nka,Sreenath:2014nca}, 
until very recently, there has been little effort towards computing these 
quantities in multi field models. 
As we were converging on the manuscript, there appeared three coordinated efforts 
wherein the scalar bispectrum has been numerically evaluated in multi field 
models~\cite{Dias:2016rjq,Seery:2016lko,Mulryne:2016mzv}.
While these efforts are indeed more comprehensive and focus on the important 
case of scalars, the approach adopted in these efforts (the so-called transport 
method) is different from the method we work with.
Our eventual goal is to arrive at a numerical procedure to evaluate all the 
three-point functions in two field and, in general, multi field models.
In contrast to the scalars, the tensor perturbations are simpler to study as
they depend only on the evolution of the scale factor.
As a first step of the process, in this work, we compute the tensor bispectrum
and the corresponding non-Gaussianity parameter in two field models of inflation.
To check the accuracy of the numerical procedure, we first consider simple 
situations leading to slow roll inflation and compare the numerical results
with the analytical results available in such cases. 
We then study the effects of the curved trajectory in the field space on the 
tensor bispectrum and the corresponding non-Gaussianity parameter.
We also explicitly examine the validity of the consistency relation governing 
the three-point function in the squeezed limit and discuss the contributions 
to tensor non-Gaussianities during the epoch of preheating.

\par

This paper is organized as follows. 
In the following section, we shall quickly summarize the equations of motion 
describing the background dynamics of inflationary scenarios driven by two 
canonical scalar fields. 
In Sec.~\ref{sec:tps}, we shall outline the quantization of the tensor modes 
and the definition of the tensor power spectrum. 
In Sec.~\ref{sec:tbs-hnl}, we shall present the essential expressions governing
the tensor bispectrum arrived at using the Maldacena formalism and introduce
the dimensionless non-Gaussianity parameter $\hnl$ that characterizes the 
amplitude of the tensor bispectrum.
In Sec.~\ref{sec:tbs-hnl-ds}, we shall discuss the analytical results for the
tensor bispectrum and the corresponding non-Gaussianity parameter in the de 
Sitter limit.
In Sec.~\ref{sec:ne}, we shall describe the numerical procedure that we 
adopt to calculate the tensor bispectrum and the non-Gaussianity parameter and 
then go to on to evaluate these quantities in three different two field models, 
\viz\/ double inflation, hybrid inflation and aligned natural inflation. 
Moreover, in the case of double inflation, we study the imprints of turning 
trajectories on~$\hnl$.
In Sec.~\ref{sec:sl}, using our numerical techniques, we also examine the
so-called consistency condition relating the tensor bispectrum to the 
tensor power spectrum in the squeezed limit, wherein one of the wavenumbers 
involved is much smaller than the other two.
In Sec.~\ref{sec:cdp}, we shall discuss the effects of preheating on the 
non-Gaussianity parameter $\hnl$. 
Lastly, in Sec.~\ref{sec:d}, we shall conclude with a brief summary. 

\par

Note that, we shall work with natural units wherein $\hbar=c=1$, and define 
the Planck mass to be $\Mpl=(8\, \pi\, G)^{-1/2}$. 
We shall adopt the metric signature of $(-,+,+,+)$.
As usual, overdots and overprimes shall denote differentiation with respect
to the cosmic and the conformal time coordinates, respectively.
Also, $N$ shall refer to the number of e-folds.

%%%%%%%%%%%%%%%%%%%%%%%%%%%%%%%%%%%%%%%%%%%%%%%%%%%%%%%%%%%%%%%%%%%%%%%%%%%%%%%

\section{Background equations}

We shall consider the background to be the spatially flat, 
Friedmann-Lema\^itre-Robertson-Walker (FLRW) metric that is described by the 
line-element
\begin{equation}
\d s^2=-\d t^2+a^2(t)\,\delta_{ij}\,\d x^i\, \d x^j
=a^2(\eta)\,\l(-\d \eta^2+\delta_{ij}\,\d x^i\, \d x^j\r),
\end{equation}
where the quantity $a$ denotes the scale factor, while $t$ and $\eta=\int \d t/
a(t)$ represent the cosmic and the conformal time coordinates. 
We shall study inflationary models consisting of two scalar fields, say, $\phi$ 
and $\chi$, that are described by the action
\begin{equation}
{\cal S}[\phi_I] 
= \int \d^4 x\, \sqrt{- g}\, 
\l[-\frac{1}{2}\,\sum_{I=1}^2 \partial_{\mu}\phi_I\,
\partial^{\mu}\phi_I - V(\phi_I)\r],\label{eq:S}	
\end{equation}
where $\phi_I=\{\phi,\chi\}$ and $V(\phi_I)$ is the potential characterizing 
the scalar fields.
The equations of motion that govern the homogeneous components of these scalar 
fields are given by
\begin{equation}
{\ddot \phi_I}+3\, H\, {\dot \phi_I} + V_{I}=0,\label{eq:phichi}
\end{equation}
where $V_{I}=\partial V/\partial \phi_I$.
The quantity $H= \dot{a}/a$ denotes the Hubble parameter and its evolution 
is described by the following Friedmann equation:
\begin{eqnarray}
H^2= \frac{1}{3\,\Mp^2}\,
\l[\frac{1}{2}\, \sum_{I=1}^2 {\dot \phi_I}^2+V(\phi_I)\r].\label{eq:ffe}
\end{eqnarray}
It is useful to introduce here the so-called first slow roll parameter
$\epsilon_1$, which is defined as
\begin{equation}
\epsilon_1=-\f{\dot{H}}{H^2}.
\end{equation}

%%%%%%%%%%%%%%%%%%%%%%%%%%%%%%%%%%%%%%%%%%%%%%%%%%%%%%%%%%%%%%%%%%%%%%%%%%%%%%%

\section{The tensor modes and the power spectrum}\label{sec:tps}

As we have mentioned earlier, we shall be focusing on the tensor perturbations
in this work. 
When the tensor perturbations are taken into account, the FLRW metric can be 
expressed as~\cite{Maldacena:2002vr}
\begin{equation}
\d s^2 =a^{2}(\eta)\, \l\{-\d \eta^2 
+ \l[{\rm e}^{\gamma(\eta,{\bm x})}\r]_{ij}\,
\d x^i\, \d x^j\r\},\label{eq:metric}
\end{equation}
where $\gamma_{ij}$ is a symmetric, transverse and traceless tensor.
At the quadratic order, the action governing the tensor perturbations is given 
by~\cite{Mukhanov:1990me,Maldacena:2002vr}
\begin{equation}
{\cal S}_{2}[\gamma_{ij}]
=\f{\Mpl^2}{8}\, \int \d \eta\; \int \d^{3}{\bm x}\;\, a^{2}\,
\l[{\gamma_{ij}'}^{\!\!2}-\l(\pa\gamma_{ij}\r)^{2}\r],\label{eq:a-gg}
\end{equation}
which, evidently, leads to a linear equation of motion.
In Fourier space, the tensor modes, say, $h_k$, are found to satisfy the 
differential equation
\begin{equation}
h_k''+2\, \f{a'}{a}\, h_k' + k^2\, h_k= 0.\label{eq:de-hk}
\end{equation}
On quantization, the tensor perturbation $\g_{ij}$ can be decomposed in terms 
of the Fourier modes $h_k$ as follows:
\begin{eqnarray}
{\hat \gamma}_{ij}(\eta, {\bm x}) 
&=& \int \frac{\d^{3}{\bm k}}{\l(2\,\pi\r)^{3/2}}\,
\hat{\gamma}_{ij}^{\bm k}(\eta)\, {\rm e}^{i\,{\bm k}\cdot{\bm x}}\nn\\
&=& \sum_{s}\int \frac{\d^{3}{\bm k}}{(2\,\pi)^{3/2}}\,
\l[\hat{a}^{s}_{\bm k}\, \varepsilon^{s}_{ij}({\bm k})\,
h_{k}(\eta)\, {\rm e}^{i\,{\bm k}\cdot{\bm x}}
+\hat{a}^{s\dagger}_{\bf k}\,\varepsilon^{s*}_{ij}({\bm k})\, h^{*}_{k}(\eta)\,
{\rm e}^{-i\,{\bm k}\cdot{\bm x}}\r],\label{eq:t-m-dc}
\end{eqnarray}
where the annihilation operators $\hat{a}_{\bm k}^{s}$ and the creation operators
$\hat{a}^{s\dagger}_{\bm k}$ satisfy the standard commutation relations.
The quantity $\varepsilon^{s}_{ij}({\bm k})$ represents the polarization 
tensor of the gravitational waves with their helicity being denoted by the 
index~$s$.
The transverse and traceless nature of the gravitational waves lead to
the conditions $k_{i}\,\varepsilon_{ij}^s({\bm k})=\varepsilon^{s}_{ii}({\bm k})=0$. 
We shall choose to work with the following normalization of the polarization tensor:
$\varepsilon_{ij}^{r}({\bm k})\, \varepsilon_{ij}^{s*}({\bm k})
=2\,\delta^{rs}$~\cite{Maldacena:2002vr}.

\par

It is often convenient to rewrite the modes $h_k$ in terms of the 
corresponding Mukhanov-Sasaki variable $u_k=\Mpl\, a\,h_k/\sqrt{2}$.
Then, the Mukhanov-Sasaki variable $u_k$ satisfies the equation 
\begin{equation}
u_k''+\l(k^2-\f{a''}{a}\r)\,u_k=0.\label{eq:de-uk} 
\end{equation}
It is useful to note that the quantity $a''/a$ can be expressed in terms 
of the slow roll parameter $\epsilon_1$ as
\begin{equation}
\f{a''}{a}=(a\, H)^2\, \l(2-\epsilon_1\r).
\end{equation} 
The initial conditions for the differential equation~(\ref{eq:de-uk}) are
imposed when the modes are well inside the Hubble radius, \ie\/ when 
$k/(a\,H)\gg 1$. 
In this sub-Hubble limit, the following positive frequency solution of the 
Mukhanov-Sasaki variable $u_k$ is chosen as the initial condition:
\begin{equation}
u_k(\eta)=\f{1}{\sqrt{2\,k}}\, {\rm e}^{-i\,k\,\eta}.\label{eq:bdin}
\end{equation}
This condition is commonly referred to as the Bunch-Davies initial condition. 

\par

The tensor power spectrum, \viz\/ ${\mathcal P}_{_{\rm T}}(k)$, evaluated at 
a suitably late conformal time, say, $\eta_{\rm e}$, is defined as 
\begin{eqnarray}
\langle\, {\hat \gamma}_{ij}^{\bm k}(\eta_{\rm e})\,
{\hat \gamma}_{mn}^{\bm k'}(\eta_{\rm e})\,\rangle
&=&\f{(2\,\pi)^2}{2\, k^3}\, \f{\Pi_{ij,mn}^{{\bm k}}}{4}\;
{\mathcal P}_{_{\rm T}}(k)\;
\delta^{(3)}({\bm k}+{\bm k'}),\label{eq:tps-d}
\end{eqnarray}
with the expectation values on the left hand side to be evaluated in the 
specified initial quantum state, and the quantity $\Pi_{ij,mn}^{\vk}$ is 
given by
\begin{equation}
\Pi_{ij,mn}^{\vk}
=\sum_{s}\;\varepsilon_{ij}^{s}(\vk)\;
\varepsilon_{mn}^{s\ast}(\vk).
\end{equation}
On making use of the decomposition~(\ref{eq:t-m-dc}), the inflationary tensor 
power spectrum evaluated in the vacuum state $\vert 0\rangle$ (such that 
$\hat{a}_{\bm k}^{s} \vert 0\rangle =0$ $\forall\;{\bm k}$ and $s$) can be 
expressed as
\begin{equation}
{\mathcal P}_{_{\rm T}}(k)
= 4\;\f{k^3}{2\, \pi^2}\, \vert h_k\vert^2.\label{eq:tps}
\end{equation}
The amplitude $\vert h_k\vert$ on the right hand side of this expression is to 
be evaluated when the modes are sufficiently outside the Hubble radius, \ie\/ 
when $k/(a\,H)\ll1 $.
It should be mentioned here that the tensor spectral index $\nt$ is defined 
as 
\begin{equation}
\nt = \f{\d \ln \pt(k)}{\d \ln k}.\label{eq:nt} 
\end{equation}

%%%%%%%%%%%%%%%%%%%%%%%%%%%%%%%%%%%%%%%%%%%%%%%%%%%%%%%%%%%%%%%%%%%%%%%%%%%%%%%%

\section{The tensor bispectrum and the corresponding
non-Gaussianity\\ parameter}\label{sec:tbs-hnl}

The dominant signatures of non-Gaussianities are the three-point functions. 
The tensor bispectrum, \viz\/ the three-point correlation function 
describing the tensor perturbations, that arises in a given inflationary
model can be evaluated using the so-called Maldacena 
formalism~\cite{Maldacena:2002vr}.
The formalism involves first deriving the cubic order action governing the  
perturbations.
At the cubic order, the action describing the tensor perturbations is found
to be
\begin{equation} 
S_3[\gamma_{ij}] 
= \frac{\Mp^{2}}{2}\,\int\d\eta\,\int \d^{3}\bm{x}\; 
\biggl[\frac{a^2}{2}\,\gamma_{lj}\,\gamma_{im}\,\pa_l\pa_m\gamma_{ij}
-\,\frac{a^2}{4}\,\gamma_{ij}\,\gamma_{lm}\,
\pa_l\pa_m\gamma_{ij}\biggr].\label{eq:a-ttt}
\end{equation}
Given this action, the corresponding three-point function can then be
arrived at using the standard techniques of quantum field theory.
In this section, we shall gather the essential expressions describing the
tensor bispectrum.
We shall also define the corresponding dimensionless non-Gaussianity 
parameter that can be introduced for conveniently characterizing the 
amplitude of the tensor bispectrum, as is popularly done in the scalar
case~\cite{Sreenath:2013xra}.

\par

The tensor bispectrum in Fourier space, \viz\/
$\cB_{\gamma\gamma\gamma}^{m_3n_3}(\vka,\vkb,\vkc)$, evaluated towards the 
end of inflation at the conformal time, say, $\ee$, is defined as
\begin{eqnarray}
\langle\, {\hat \gamma}_{m_1n_1}^{\vka}(\eta _{\rm e})\,
{\hat \gamma}_{m_2n_2}^{\vkb}(\eta _{\rm e})\, 
{\hat \gamma}_{m_3n_3}^{\vkc}(\eta _{\rm e})\,\rangle 
&=& \l(2\,\pi\r)^3\, 
\cB_{\g\g\g}^{m_1n_1m_2 n_2 m_3 n_3}(\vka,\vkb,\vkc)\;
\delta^{(3)}\l(\vka+\vkb+\vkc\r).\label{eq:ttt}\nn\\
\end{eqnarray}
It should be mentioned that the delta function on the right hand side implies 
that the wavevectors $\vka$, $\vkb$ and $\vkc$ form the edges of a triangle.
For convenience, hereafter, we shall set
\begin{equation}
\cB_{\g\g\g}^{m_1n_1m_2n_2m_3n_3}(\vka,\vkb,\vkc)
= \l(2\,\pi\r)^{-9/2}\, G_{\g\g\g}^{m_1n_1m_2n_2m_3n_3}(\vka,\vkb,\vkc).
\end{equation}
The quantity $G_{\gamma\gamma\gamma}^{m_1n_1m_2n_2m_3n_3}(\vka,\vkb,\vkc)$,
evaluated in the perturbative vacuum, can be obtained to be (see, for 
instance, Refs.~\cite{Maldacena:2002vr,Sreenath:2013xra})
\begin{eqnarray} \label{eq:Gggg}
G_{\gamma\gamma\gamma}^{m_1n_1m_2n_2m_3n_3}(\vka,\vkb,\vkc)
&= & \Mp^2\; \biggl[\bigl(\Pi_{m_1n_1,ij}^{\vka}\,\Pi_{m_2n_2,im}^{\vkb}\,
\Pi_{m_3n_3,lj}^{\vkc}\nn\\
& &-\,\f{1}{2}\;\Pi_{m_1n_1,ij}^{\vka}\,\Pi_{m_2n_2,ml}^{\vkb}\,
\Pi_{m_3n_3,ij}^{\vkc}\bigr)\, k_{1m}\, k_{1l}
+{\rm five~permutations}\biggr]\nn\\
& &\times\; \bigl[h_{\ska}(\ee)\, h_{\skb}(\ee)\, h_{\skc}(\ee)\,
\cG_{\gamma\gamma\gamma}(\vka,\vkb,\vkc)\nn\\
& &+\, {\rm complex~conjugate}\bigr],\label{eq:tbs-ov}
\end{eqnarray}
where $\cG_{\gamma\gamma\gamma}(\vka,\vkb,\vkc)$ is described by the integral
\begin{equation}
\cG_{\g\g\g}(\vka,\vkb,\vkc)
=-\f{i}{4}\,\int_{\ei}^{\ee} \d\eta\; a^2\, h_{\ska}^{\ast}\,
h_{\skb}^{\ast}\,h_{\skc}^{\ast},\label{eq:cGggg}
\end{equation}
with $\ei$ denoting the time when the initial conditions are imposed on the 
perturbations.

\par

As is well known, in the case of the scalars, a dimensionless non-Gaussianity 
parameter is often introduced (in fact, a set of three parameters are 
considered) to roughly characterize the amplitude of the scalar bispectrum. 
A similar dimensionless quantity can be introduced to describe the tensor 
bispectrum.
It can be defined to be the following dimensionless ratio of the tensor 
bispectrum and the power spectrum~\cite{Sreenath:2013xra}:
\begin{eqnarray}
\hnl(\vka,\vkb,\vkc)
&=&-\l(\f{4}{2\,\pi^2}\r)^2\,
\l[k_1^3\, k_2^3\, k_3^3\;
G_{\g\g\g}^{m_1n_1m_2n_2m_3n_3}(\vka,\vkb,\vkc)\r]\nn\\
&\times&\; \l[\Pi_{m_1n_1,m_3n_3}^{\vka}\,\Pi_{m_2n_2,{\bar m}{\bar n}}^{\vkb}\;
k_3^3\; {\mathcal P}_{_{\rm T}}(k_1)\;{\mathcal P}_{_{\rm T}}(k_2)
+{\rm five~permutations}\r]^{-1},\qquad\label{eq:hnl-ov}
\end{eqnarray}
where the overbars on the indices imply that they need to be summed over all
allowed values.
Since we shall be focusing here only on the amplitude of the tensor bispectrum, 
for simplicity, we shall set the polarization tensor to unity.
In such a case, the expression~(\ref{eq:tbs-ov}) for the tensor bispectrum 
reduces to
\begin{eqnarray}
G_{\gamma\gamma\gamma}(\vka,\vkb,\vkc)
&=& \Mp^2\,\bigl[h_{\ska}(\ee)\, h_{\skb}(\ee)\,h_{\skc}(\ee)\,
{\bar \cG}_{\gamma\gamma\gamma}(\vka,\vkb,\vkc)\nn\\
& &+\, {\rm complex~conjugate}\bigr],\label{eq:tbs}
\end{eqnarray}
where the quantity ${\bar \cG}_{\g\g\g}(\vka,\vkb,\vkc)$ is described by
the integral
\begin{eqnarray}
{\bar \cG}_{\g\g\g}(\vka,\vkb,\vkc)
&=&-\f{i}{4}\,\l(k_1^2+k_2^2+k_3^2\r)\,\int_{\ei}^{\ee} \d\eta\; 
a^2\, h_{\ska}^{\ast}\,
h_{\skb}^{\ast}\,h_{\skc}^{\ast}.\label{eq:cGggg-fv}
\end{eqnarray}
Also, if we ignore the factors involving the polarization tensor, the 
non-Gaussianity parameter $\hnl$ simplifies to
\begin{eqnarray}
\hnl(\vka,\vkb,\vkc)
&=&-\l(\f{4}{2\,\pi^2}\r)^2\, \l[k_1^3\, k_2^3\, k_3^3\; 
G_{\g\g\g}(\vka,\vkb,\vkc)\r]\nn\\ 
& &\times\,\biggl[2\,k_3^3\; {\mathcal P}_{_{\rm T}}(k_1)\,
{\mathcal P}_{_{\rm T}}(k_2)+\,{\rm two~permutations}\biggr]^{-1}.
\label{eq:hnl}
\end{eqnarray}

%%%%%%%%%%%%%%%%%%%%%%%%%%%%%%%%%%%%%%%%%%%%%%%%%%%%%%%%%%%%%%%%%%%%%%%%%%%%%%%

\section{The non-Gaussianity parameter $\hnl$ in slow roll 
inflation}\label{sec:tbs-hnl-ds}

Evidently, in order to evaluate the tensor bispectrum, we shall first
require the modes~$h_k$.
Also, using the modes, we need to be able to evaluate the 
integral~(\ref{eq:cGggg-fv}) and the asymptotic forms of the modes to
arrive at the tensor bispectrum and the corresponding non-Gaussianity
parameter.
In slow roll inflation, one often works in the de Sitter approximation
wherein the tensor modes~$h_k$ are given by
\begin{equation}
h_k(\eta) = \f{\sqrt{2}\,i\,H_0}{\Mp\,\sqrt{2\,k^3}}\,
\l(1+i\,k\,\eta\r)\, {\rm e}^{-i\,k\,\eta},\label{eq:hk-ds}
\end{equation}
with $H_0$ being the Hubble parameter in de Sitter inflation.
These modes can be easily used to arrive at the following well-known, 
strictly scale invariant tensor power spectrum (evaluated towards the
end of inflation, \ie\/ as $\ee\to 0$):
\begin{equation}
\pt(k)=\f{2\, H_0^2}{\pi^2\, \Mp^2}.
\end{equation}

\par

Using the modes~(\ref{eq:hk-ds}), the integral~(\ref{eq:cGggg-fv}) can 
be evaluated to be
\begin{equation}
{\bar \cG}_{\g\g\g}(\vka,\vkb,\vkc)
=-\f{i\, H_0\, \l(k_1^2+k_2^2+k_3^2\r)}{4\,\Mpl^3\, \l(k_1\,k_2\,k_3\r)^{3/2}}\,
\l(k_{_{\rm T}} -\f{k_1\, k_2+k_1\, k_3 + k_2\, k_3}{k_{_{\rm T}}}
-\f{k_1\, k_2\, k_3}{k_{_{\rm T}}^2}\r),
\end{equation}
where $k_{_{\rm T}}=k_1+k_2+k_3$.
In the limit $\ee\to 0$, the corresponding tensor bispectrum 
$G_{\gamma\gamma\gamma}(\vka,\vkb,\vkc)$ and the non-Gaussianity
parameter $\hnl(\vka,\vkb,\vkc)$ can be obtained to be
\begin{equation}
G_{\gamma\gamma\gamma}(\vka,\vkb,\vkc)
= -\f{H_0^4\, \l(k_1^2+k_2^2+k_3^2\r)}{2\,\Mpl^4\, \l(k_1\,k_2\,k_3\r)^3}\,
\l(k_{_{\rm T}} -\f{k_1\, k_2+k_1\, k_3 + k_2\, k_3}{k_{_{\rm T}}}
-\f{k_1\, k_2\, k_3}{k_{_{\rm T}}^2}\r)
\end{equation}
and
\begin{equation}
\hnl(\vka,\vkb,\vkc)
=\f{1}{4}\,\l(\f{k_1^2+k_2^2+k_3^2}{k_1^3+k_2^3+k_3^3}\r)\,
\l(k_{_{\rm T}}
-\f{k_1\, k_2+k_2\, k_3 + k_3\, k_1}{k_{_{\rm T}}}
-\f{k_1\, k_2\, k_3}{k_{_{\rm T}}^2}\r).
\end{equation}
Note that, in the equilateral limit (\ie\/ when $k_1=k_2=k_3$), we have 
$\hnl=17/36 \simeq 0.472$, while in the squeezed limit  (\ie\/ as $k_1=
k_2$ and $k_3\to0$), we have $\hnl = 3/8 = 0.375$.
These analytical results prove to be very handy for examining the
accuracy of the numerical procedures that we shall adopt to
evaluate the tensor modes, the tensor power spectrum and the
tensor bispectrum. 

%%%%%%%%%%%%%%%%%%%%%%%%%%%%%%%%%%%%%%%%%%%%%%%%%%%%%%%%%%%%%%%%%%%%%%%%%%%%%%%%

\section{Numerical evaluation}\label{sec:ne}

As we have described before, our aim in this work is to numerically evaluate the 
magnitude and shape of the non-Gaussianity parameter $\hnl$ in two field models 
of inflation.
We shall make use of the analytical results available in the slow roll limit 
(actually, in the de Sitter limit) to check the accuracy of our numerical 
results.  
In this section, we shall first quickly describe the numerical procedure that we 
shall adopt for the evaluation of the non-Gaussianity parameter $\hnl$. 
Thereafter, we shall consider three specific inflationary models and evaluate the 
non-Gaussianity parameter $\hnl$ in these models.

\par

Evidently, we shall first require the behavior of the background quantities 
and the tensor modes.
Once these are at hand, the tensor bispectrum~(\ref{eq:tbs}) can be 
arrived at by computing the integral~(\ref{eq:cGggg-fv}) and then
using the asymptotic forms of the tensor modes. 
These quantities can be utilized to finally obtain the non-Gaussianity
parameter $\hnl$.

\par

Our numerical procedure is essentially similar to an earlier work in this
context which had dealt with single field models of 
inflation~\cite{Sreenath:2013xra}. 
Once the parameters in the potential and the initial conditions are specified, 
one can integrate the equations~(\ref{eq:phichi}) that govern the scalar
fields and the Friedmann equation~(\ref{eq:ffe}) to arrive at the evolution 
of the background quantities. 
Usually the initial value of the fields are chosen to lead to enough number 
of e-folds (say, $60$-$70$ e-folds of inflation). 
Once we have the background quantities, we can solve for the tensor 
perturbations by integrating the governing equation~(\ref{eq:de-hk}),
along with the Bunch-Davies initial condition~(\ref{eq:bdin}). 
In this computation, the initial conditions are imposed on the modes when they 
are sufficiently inside the Hubble radius [we have chosen when $k/(a\,H) =10^2$]. 
The power spectrum is evaluated in the super-Hubble domain, when the amplitude 
of the modes have reached a constant value [which typically occur when $k/(a\,H)
\simeq 10^{-5}$]. 

\par

We solve the background and the perturbation equations as functions of the 
number of e-folds  using the fifth order Runge-Kutta algorithm (see, for 
instance, Ref.~\cite{Press:2007:NRE:1403886}).
Since the tensor mode is constant during the super Hubble evolution, we can 
neglect the contribution of $\hnl$ during this period (for more details, see
Ref.~\cite{Sreenath:2013xra}). 
This simplifies the numerical integration involved in the calculation of 
$\hnl$. 
Note that the modes oscillate strongly in the sub-Hubble domain, leading
to oscillating integrands. 
In order to handle such integrands, an exponential cut-off is included to
regulate the integrals in the sub-Hubble domain, as has been implemented 
earlier in similar contexts (in this context, see Refs.~\cite{Sreenath:2013xra,
Hazra:2012yn,Chen:2006xjb}).
Such a cut-off can be justified theoretically as it helps in identifying the
correct perturbative vacuum~\cite{Hazra:2012yn,Seery:2005wm,Chen:2008wn}.
The integration is carried out using the Bode’s rule\footnote{We should add 
that there is some confusion concerning whether it is Bode's or Boole's 
rule~\cite{Press:2007:NRE:1403886}.}, from the earliest time $\eta_i$ when 
the smallest of the three wavenumbers $(k_1, k_2, k_3)$ is well inside the 
Hubble radius to the final time $\eta_e$ when the largest of them is 
sufficiently outside the Hubble radius.

%%%%%%%%%%%%%%%%%%%%%%%%%%%%%%%%%%%%%%%%%%%%%%%%%%%%%%%%%%%%%%%%%%%%%%%%%%%%%%%%

\subsection{Double inflation}

The simplest of two field inflationary models is the model which is described by 
the potential~\cite{Silk:1986vc,Holman:1991ht}
\begin{equation}
V(\phi, \chi)
=\f{1}{2}\,m_{\phi}^2\, \phi^2+ \f{1}{2}\,m_{\chi}^2\, \chi^2.\label{eq:double}
\end{equation}
This model is often referred to as double inflation, since it can lead to two 
different epochs of inflation (characterized by different values of the 
Hubble parameter) if the parameters $m_{\phi}$ and $m_{\chi}$ are very different.
Even though this model seems to be ruled out by the current observations, it is 
instructive to work with this model since it is very simple. 
As we have mentioned earlier, one of our aims is to study the effects of curved 
trajectories in field space on $\hnl$ and, in this model, it is easy to construct 
different types of curved trajectories. 

\par

\begin{figure}[!htb]
\begin{center}
\includegraphics[width=7.6cm]{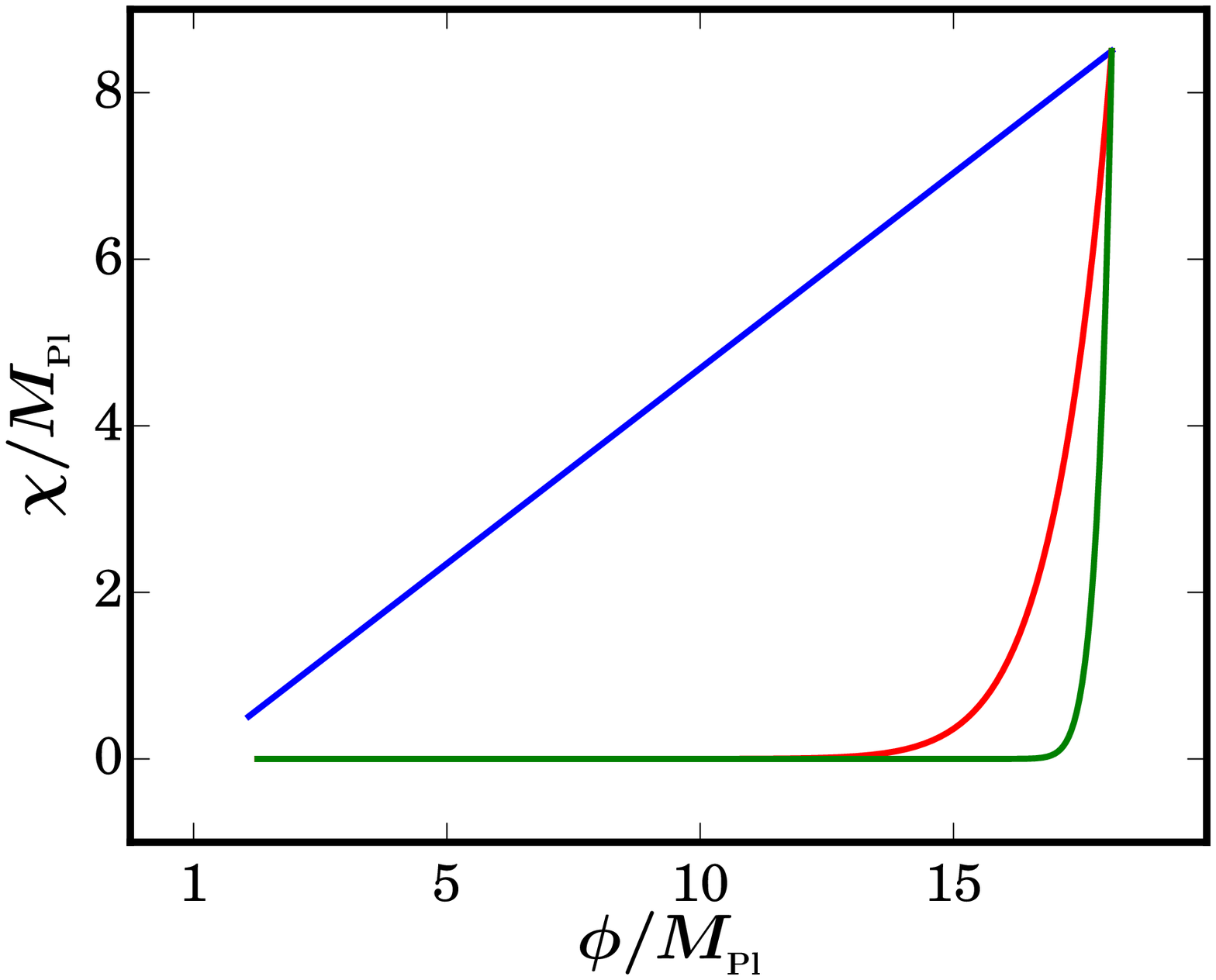} 
\includegraphics[width=7.6cm]{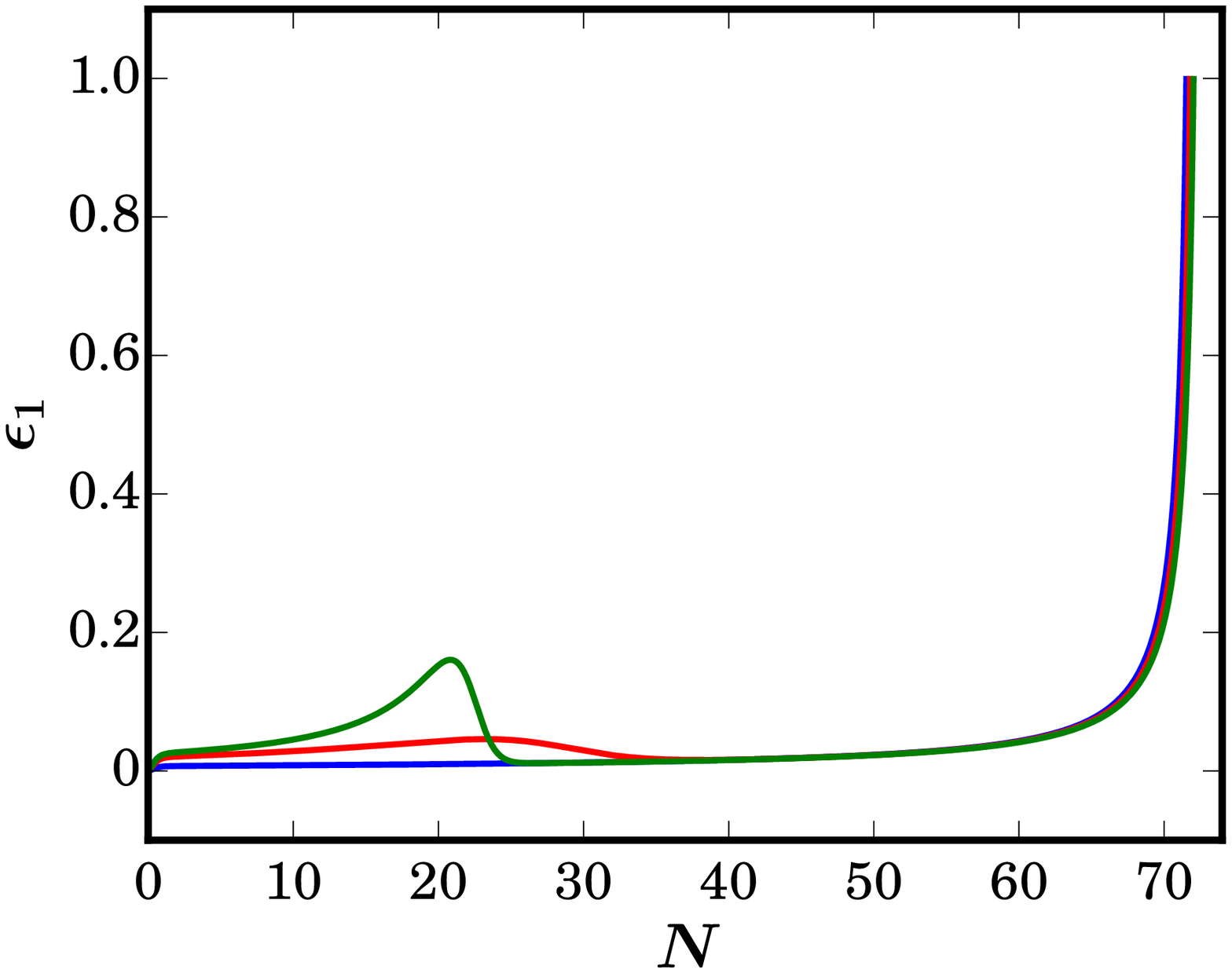}
\caption{\label{fig:di-be}
The trajectories of the fields $\phi$ and $\chi$ in the field space have been 
plotted (on the left) in the case of double inflation for different mass ratios 
$m_{\chi}=m_{\phi}$ (in blue), $m_{\chi}=4\, m_{\phi}$ (in red) and 
$m_{\chi}=8\, m_{\phi}$ (in green). 
The corresponding evolution of first slow roll parameter $\epsilon_1$ has been 
plotted as a function of the number of e-folds (on the right) with the same 
choices of colors for the different cases (as in the figure on the left).}
\end{center}
\end{figure}

For numerical analysis, we shall set $m_{\phi}=7.12 \times 10^{-6}\,\Mpl$,
and choose $m_{\chi}$ to be a multiple of $m_{\phi}$.
We shall choose the initial value of the fields to be $\phi_i=14.4\,\Mpl$ and 
$\chi_i=8.5\, \Mpl$. 
The corresponding initial velocities of the fields are chosen such that the 
first slow roll parameter $\epsilon_1$ is small.
In Fig.~\ref{fig:di-be}, we have shown the trajectories of the two scalar 
fields and the evolution of the slow roll parameter $\epsilon_1$ for 
three different mass ratios $m_{\chi}/m_{\phi}$.
%%%%%%%%%%%%%%%%%%%%%%%%%%%%%%%%%%%%%%%%%%%%%%%%%%%%%%%%%%%%%%%%%%%%%%%%%%%%%%%%
%%%%%%%%%%%%%%%%%%%%%%%%%%%%%%%%%%%%%%%%%%%%%%%%%%%%%%%%%%%%%%%%%%%%%%%%%%%%%%%% 
For $m_{\chi}= m_{\phi}$, the slow roll parameter $\epsilon_1$ is very small 
throughout inflation, as one would have naively expected. 
For $m_{\chi}=4\, m_{\phi}$, the trajectory in the field space is characterized 
by a smooth turn from a $\chi$ dominated phase to the $\chi=0$ valley and 
inflation continues along this valley. 
In the case of $m_{\chi}=8\, m_{\phi}$, the turn is more sharp and the field 
reaches the $\chi=0$ valley faster than in the former cases. 
It is important to note the effect of turning on the evolution of the first slow
roll parameter $\epsilon_1$. 
When the mass ratio increases, the turns become sharper and the slow roll 
parameter $\epsilon_1$ changes considerably during the turn.

\par

\begin{figure}[!htb] 
\begin{center}
\includegraphics[width=7.50cm]{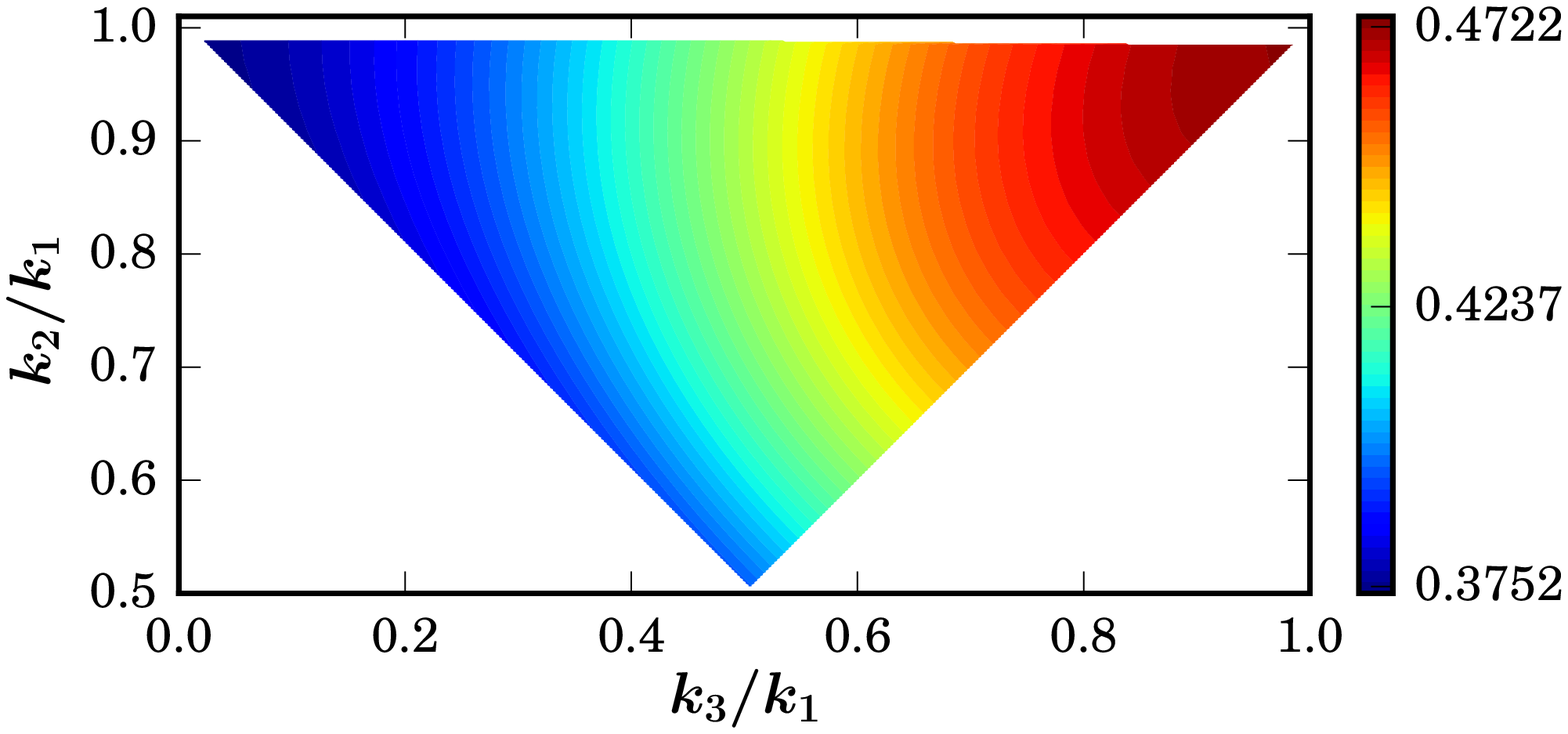}
\includegraphics[width=7.50cm]{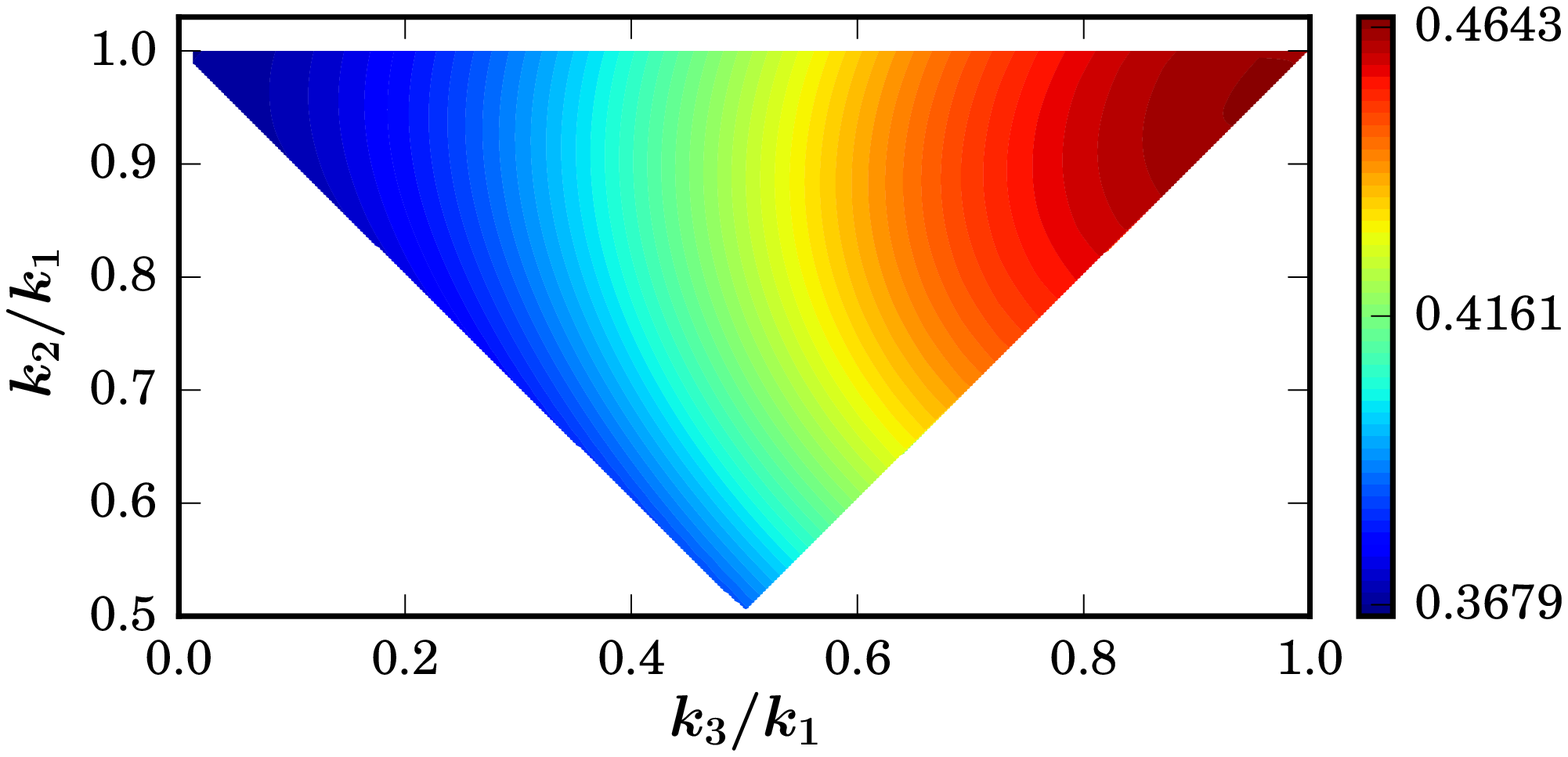}
\caption{\label{fig:hnl-c}
Density plots of $\hnl$ for an arbitrary triangular configuration of the 
wavenumbers evaluated analytically in the case of de Sitter inflation (on
the left) and obtained numerically for double inflation with $m_{\chi}=m_{\phi}$ 
(on the right).
It is evident that the analytical and the numerical results match quite well,
indicating the accuracy of the numerical procedures that have been adopted.}
\end{center}
\end{figure}

Let us now turn to understand the behavior of the non-Gaussianity parameter in 
these situations.
Since the case of $m_{\chi}=m_{\phi}$ leads to nearly de Sitter inflation, the
numerical results for $\hnl$ from this case can be compared with the analytical
results we had discussed earlier.
Evidently, this exercise can help us determine the accuracy of our numerical 
procedure.
In Fig.~\ref{fig:hnl-c}, we have illustrated the density plots of $\hnl$ for a 
triangular configuration of the wavenumbers $(k_1,k_2,k_3)$ evaluated analytically
in the case of de Sitter inflation and the numerical results for the double 
inflation model with equal values for the masses for the two fields. 
%%%%%%%%%%%%%%%%%%%%%%%%%%%%%%%%%%%%%%%%%%%%%%%%%%%%%%%%%%%%%%%%%%%%%%%%%%%%%%%%
%%%%%%%%%%%%%%%%%%%%%%%%%%%%%%%%%%%%%%%%%%%%%%%%%%%%%%%%%%%%%%%%%%%%%%%%%%%%%%%%
To arrive at the density plots of $\hnl$, we have set $k_1=5\times 10^{-2}\,
{\rm Mpc}^{-1}$, and chosen $k_2$ and $k_3$ such that $5\times10^{-4}\, 
{\rm Mpc}^{-1}< (k_2, k_3) < 5\times10^{-2}\, {\rm Mpc}^{-1}$. 
Note that the non-Gaussianity parameter $\hnl$ has an equilateral shape, \ie\/ 
its value is the largest in the equilateral limit (wherein $k_1=k_2=k_3$). 
The equilateral shape can be attributed to the fact that the non-Gaussianities 
are essentially generated as the modes leave the Hubble radius and the 
contributions on the super-Hubble scales are insignificant. 
This figure clearly illustrates that the numerical and the analytical results 
match quite well.
In fact, we find that the maximum difference between them is less than $2\%$. 

\par

\begin{figure}[!htb] 
\begin{center}
\includegraphics[width=7.50cm]{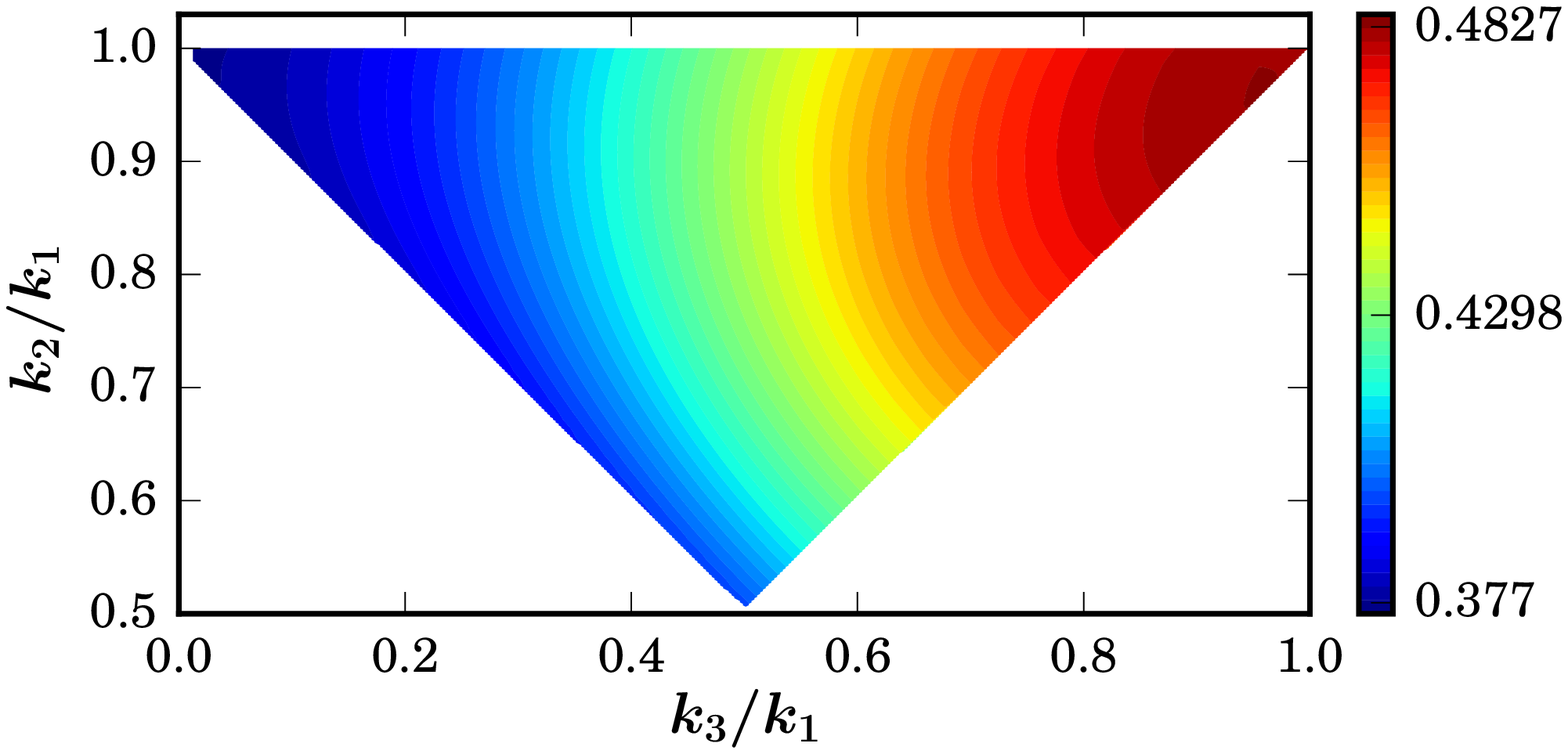} 
\includegraphics[width=7.50cm]{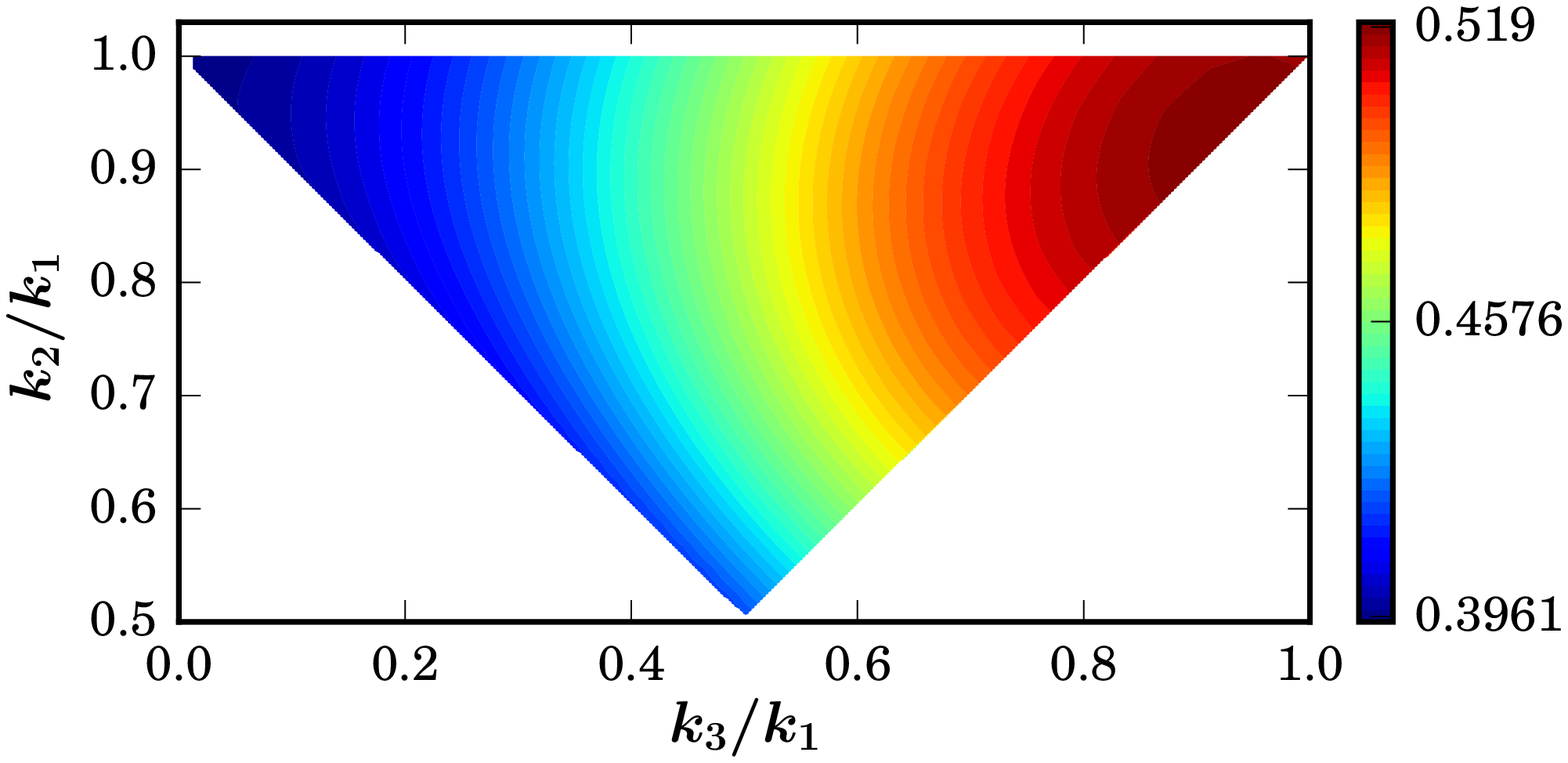}
\caption{\label{fig:hnl2}
Density plots of $\hnl$ computed numerically for an arbitrary triangular 
configuration of the wavenumbers for the case of double inflation with 
$m_{\chi}=4\,m_{\phi}$ (on the left) and $m_{\chi}=8\,m_{\phi}$ (on the
right).
Note that, in these cases, the departure from slow roll arises due to the
turn in the trajectory in the field space.
This deviation from slow roll enhances the amplitude of the non-Gaussianity
parameter $\hnl$ to some extent from the slow roll values.}
\end{center}
\end{figure}

Our next task is to study the effect of the turning of the trajectory in the 
field space on $\hnl$, and we shall utilize the cases wherein $m_{\chi}=4\, 
m_{\phi}$ and $m_{\chi}=8\, m_{\phi}$ for this purpose. 
We should mention that, in these cases, the scales of our interest leave the 
Hubble radius between the e-folds of $16$ and $33$, and the direction of the 
trajectory changes exactly in this domain.
The change in the trajectory in the field space leads to a deviation from slow 
roll, as is evident from Fig.~\ref{fig:di-be}.
This effects the tensor modes and the associated non-Gaussianities. 
In Fig.~\ref{fig:hnl2}, we have plotted the non-Gaussianity parameter $\hnl$ 
that arises in these two cases.
While the deviation from slow roll inflation clearly modifies the amplitude of 
the parameter $\hnl$, the broad equilateral shape is indeed retained. 
The departure from slow roll boosts the amplitude of $\hnl$ to a slight extent
from the slow roll values.
%%%%%%%%%%%%%%%%%%%%%%%%%%%%%%%%%%%%%%%%%%%%%%%%%%%%%%%%%%%%%%%%%%%%%%%%%%%%%%%%
%%%%%%%%%%%%%%%%%%%%%%%%%%%%%%%%%%%%%%%%%%%%%%%%%%%%%%%%%%%%%%%%%%%%%%%%%%%%%%%%
As we had mentioned, we have considered the double inflation model 
because of its simplicity.
In Fig.~\ref{fig:ps}, we have plotted the scalar (\ie\/ the adiabatic) 
and the tensor power spectra that arise in these cases.
Even the simpler case of $m_\phi=m_\chi$ will not be favored by the CMB 
data because of the large tensor-to-scalar ratio that the model leads to. 
(Recall that the tensor-to-scalar ratio $r\lesssim 0.1$, according to the 
recent Planck data~\cite{Ade:2015lrj}.)
The other two cases lead to a broad step-like feature in the power spectra.
They also result in higher scalar power on large scales and a large 
tensor-to-scalar ratio.
Due to these reasons, these cases are ruled out by the data as well.

\par

In what follows, we shall discuss two more models, \viz\/ hybrid inflation 
and aligned natural inflation. 
As we shall see, in these models, for the values of the parameters that we 
work with, the first slow roll parameter evolves smoothly and also remains 
very small during inflation. 
As a result, the non-Gaussianity parameter $\hnl$ in these models does 
not differ much from the case of de Sitter inflation.

%%%%%%%%%%%%%%%%%%%%%%%%%%%%%%%%%%%%%%%%%%%%%%%%%%%%%%%%%%%%%%%%%%%%%%%%%%%%%%%%

\subsection{Hybrid inflation}

In most of the models, inflation ends when the scalar field approaches 
the minimum of the potential.  
The hybrid model of inflation had been introduced as an alternative way 
of ending inflation~\cite{Linde:1993cn,Lyth:1996kt}. 
In this model, inflation does not end because the field reaches a minimum, 
but due to a phase transition which occurs at a critical point of one of 
the fields. 
This model is based on the potential
\begin{equation}
V(\phi,\chi)
= \frac{1}{2}\, m^2\, \phi^2 
+ \frac{\lambda}{4}\,\left(\chi^2-M^2\right)^2 
+ \frac{\lambda'}{2}\,\phi^2\, \chi^2,
\end{equation}
where $\lambda$ and $\lambda'$ are two positive coupling constants, while 
$m$ and $M$ represent two mass parameters. 
One finds that, in this model, a wide variety of trajectories are possible 
for different initial conditions~\cite{Clesse:2008pf,Clesse:2009ur}. 
When confined to domains of sub-Planckian values for the fields, the initial 
values which lead to sufficient amount of e-folds are found to be near the 
$\chi=0$ valley and as random points in the space of the scalar fields. 
But, it is observed that the initial conditions which give sufficiently 
long inflation can be always found in the region of super-Planckian 
values of the fields~\cite{Clesse:2008pf,Clesse:2009ur}.  
 
\par

In our analysis, we set $m=2.63\times 10^{-12}\,\Mp$, $M=4.14\times 
10^{-14}\,\Mp$ and $\lambda=\lambda'=2.75 \times 10^{-13}$.
The scalar fields start from the initial values $\phi_i=10.02\,\Mp$ 
and $\chi_i=21.05\,\Mp$.
The behavior of the slow roll parameter $\epsilon_1$ in this model is 
plotted in Fig.~\ref{fig:e1-hi-ni}, and the resulting tensor bispectrum 
is plotted in Fig.~\ref{fig:hnl-hi-ni}.

%%%%%%%%%%%%%%%%%%%%%%%%%%%%%%%%%%%%%%%%%%%%%%%%%%%%%%%%%%%%%%%%%%%%%%%%%%%%%%%

\subsection{Aligned natural inflation}

\begin{figure}[!htb]
\begin{center}
\includegraphics[width=8.25cm]{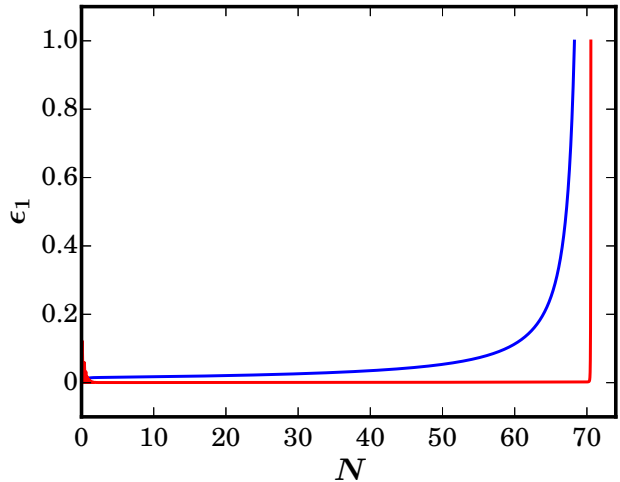} 
\caption{\label{fig:e1-hi-ni}
The evolution of first slow roll parameter $\epsilon_1$ in the case 
of hybrid inflation (in blue) and aligned natural inflation (in red).}
\end{center}
\end{figure}

The next model we shall study is the natural inflation model with a 
strong alignment~\cite{Long:2014dta,Kappl:2014lra,Peloso:2015dsa}. 
The model is described by the potential 
\begin{equation}
V(\phi,\chi)
=\Lambda^4\,\left[1-\f{1}{1+\beta}\,\cos\,(c_1\, \alpha\, \phi 
+ c_2\, \chi)
- \f{\beta}{1+\beta}\, \cos\,(c_3\, \alpha\, \phi 
+ c_4\, \chi)\r].
\end{equation}
This model also admits different types of trajectories. 
But, for the values of the parameters $\Lambda=1.76 \times 10^{-10}\,\Mp$, 
$c_1=8.20\,\Mp^{-1}$, $c_2=12.12\,\Mp^{-1}$, $c_3=8.80\,\Mp^{-1}$, 
$c_4=27.27\,\Mp^{-1}$, $\alpha=0.01$ and $\beta=0.41$, the initial 
conditions $\phi_i=24.2\,\Mp$ and $\chi_i= -0.1\,\Mp$, lead to a special 
kind of trajectory in which inflation ends due to the instability in 
the direction of the heavy field. 
This trajectory is interesting due to the fact that it leads to a 
suppressed value for the tensor-to-scalar ratio. 
The first slow roll parameter is very small throughout inflation and 
it undergoes an extremely sharp change in its value to end inflation 
(cf. Fig.~\ref{fig:e1-hi-ni}).
%%%%%%%%%%%%%%%%%%%%%%%%%%%%%%%%%%%%%%%%%%%%%%%%%%%%%%%%%%%%%%%%%%%%%%%%%%%%%%%
%%%%%%%%%%%%%%%%%%%%%%%%%%%%%%%%%%%%%%%%%%%%%%%%%%%%%%%%%%%%%%%%%%%%%%%%%%%%%%%
We shall make use of this trajectory for evaluating~$\hnl$. 

\par

\begin{figure}[!htb] 
\begin{center}
\includegraphics[width=7.50cm]{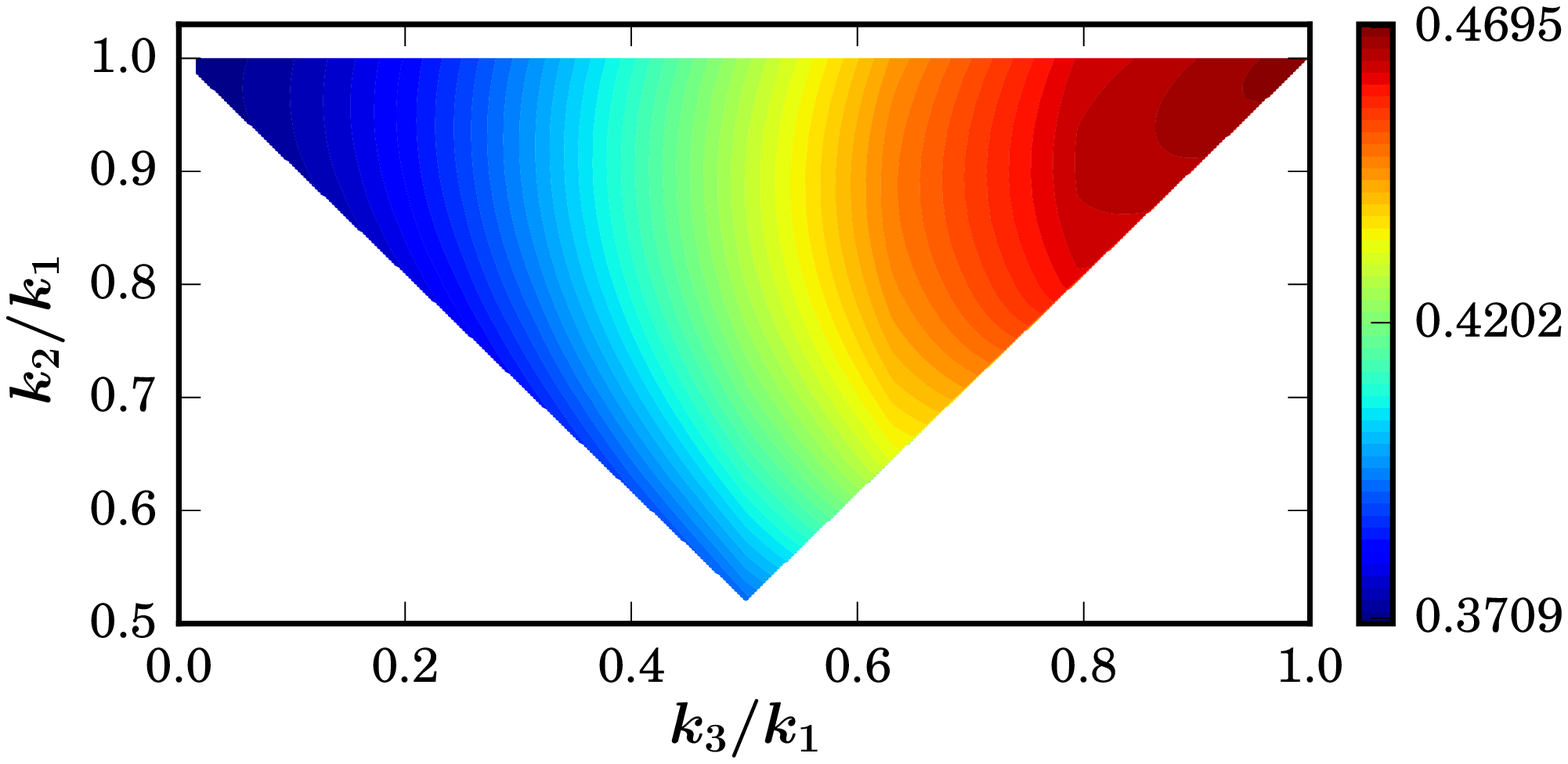} 
\includegraphics[width=7.50cm]{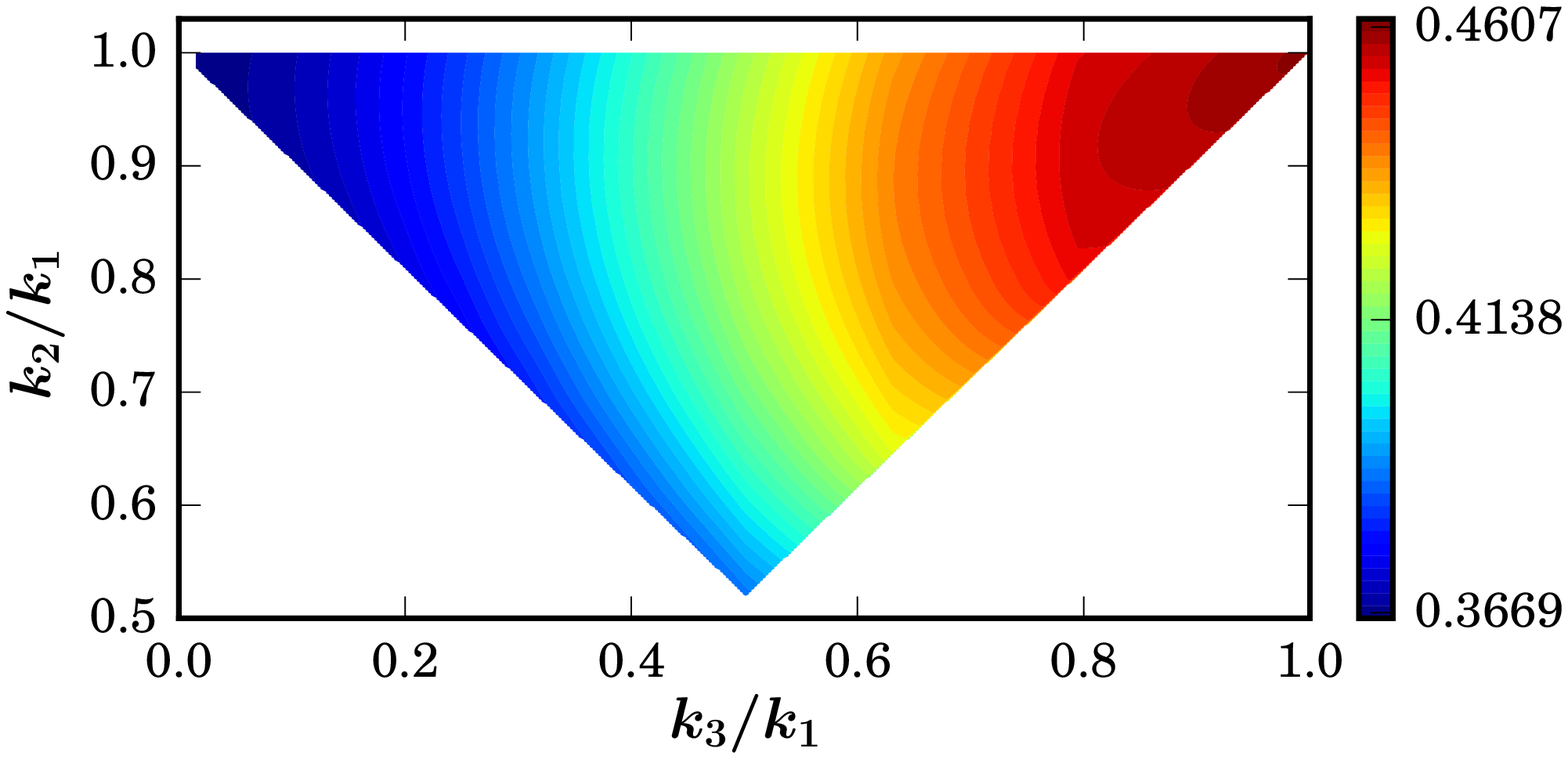} 
\caption{\label{fig:hnl-hi-ni}
Density plots of the non-Gaussianity parameter $\hnl$ evaluated numerically 
for an arbitrary triangular configuration of the wavenumbers for the case of 
hybrid inflation (on the left) and aligned natural inflation (on the right).}
\end{center}
\end{figure}

\begin{figure}[!htb]
\begin{center}
\includegraphics[width=7.50cm]{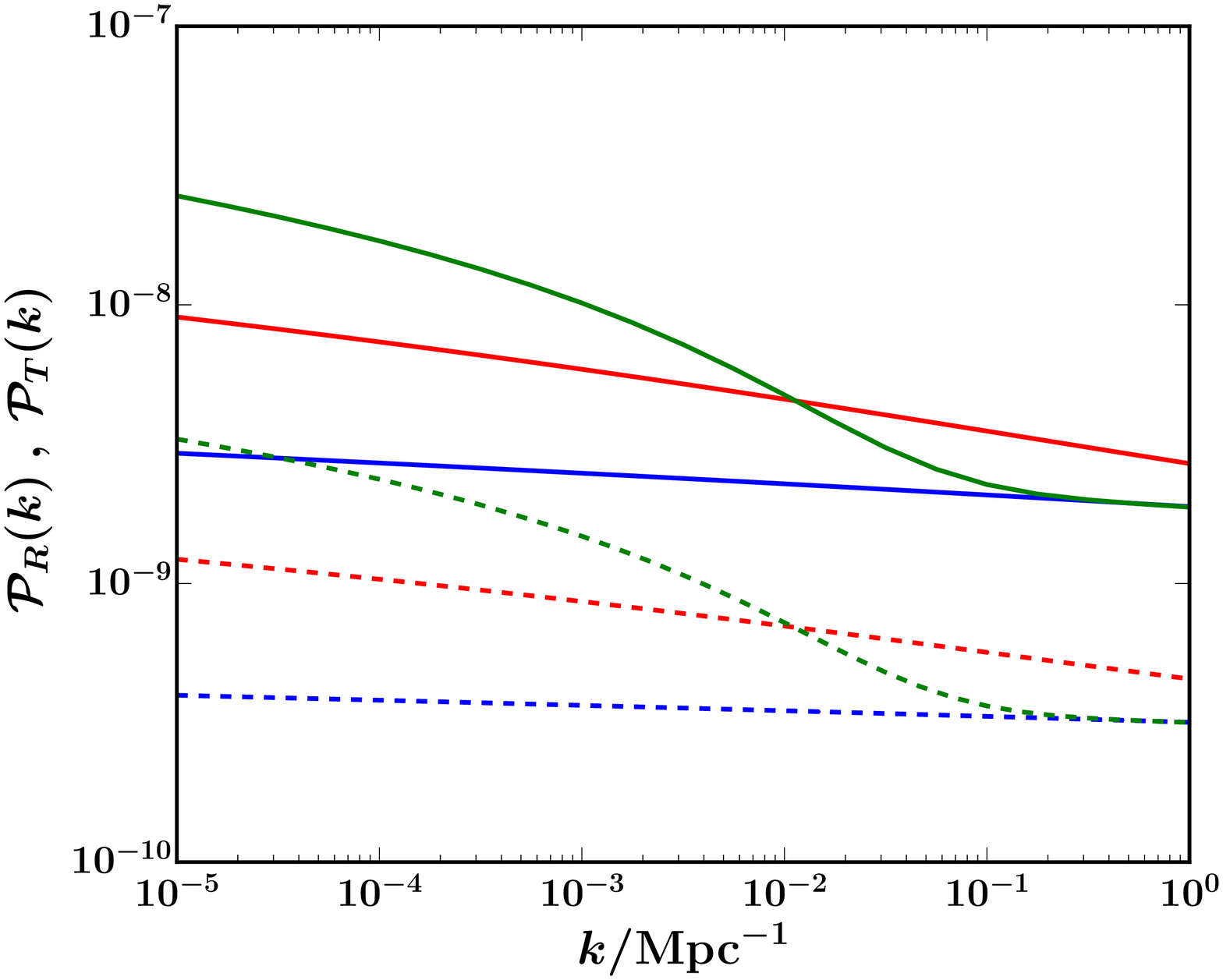}
\includegraphics[width=7.50cm]{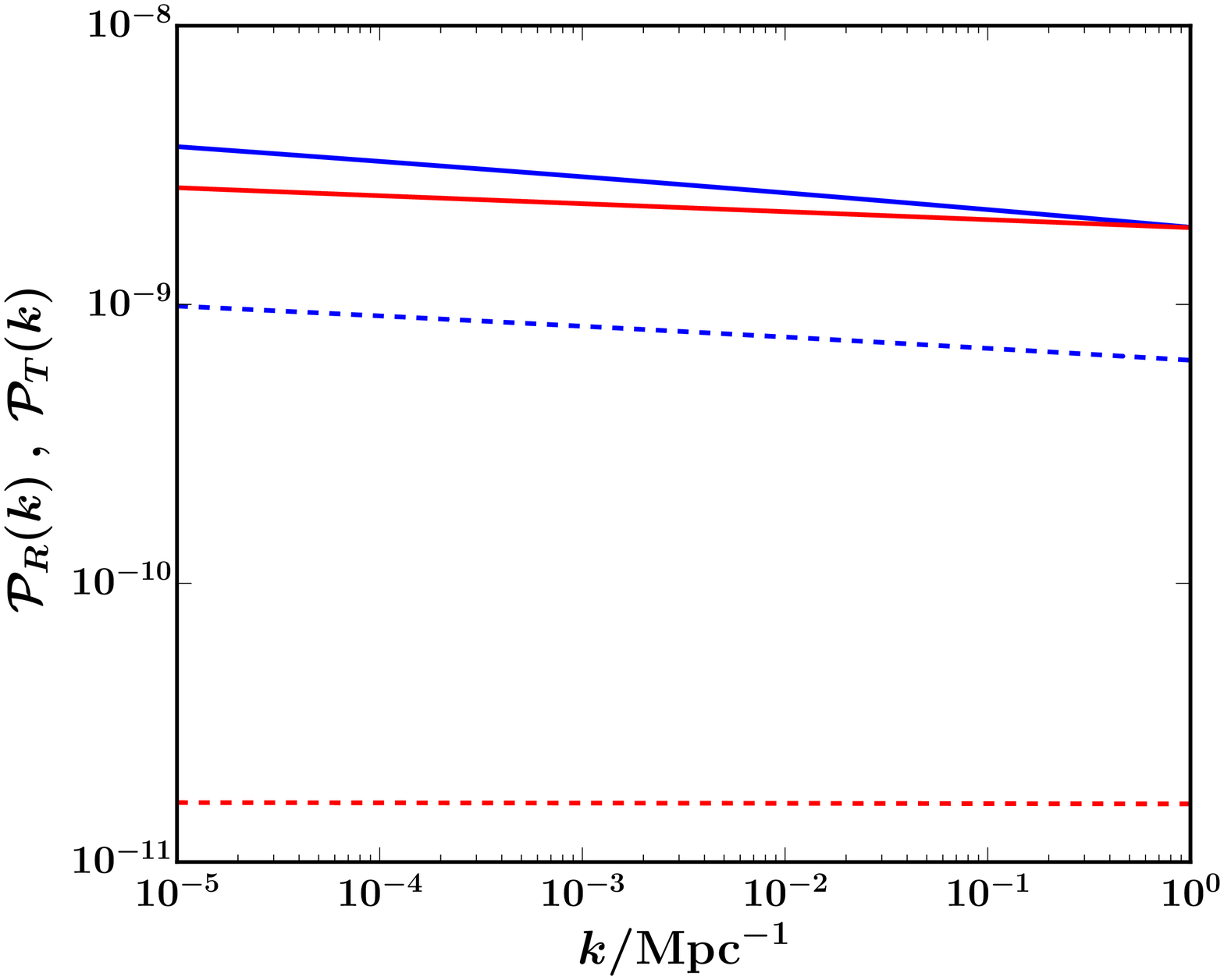}
\caption{\label{fig:ps}
The scalar, \ie\/ adiabatic (solid line) and the tensor (dashed line) power
spectra have been plotted (on the left) for the double inflation model
with $m_{\chi}=m_{\phi}$ (in blue), $m_{\chi}=4\, m_{\phi}$ (in red) and 
$m_{\chi}=8\,m_{\phi}$ (in green).
The power spectra (with same choice of lines) have also been plotted (on
the right) for the cases of the hybrid inflation (in blue) and the aligned 
natural inflation (in red) models.}
\end{center}
\end{figure}

From Fig.~\ref{fig:e1-hi-ni}, it is clear that in both the models 
(\ie\/ the  hybrid inflation model and the aligned natural inflation
model), the slow roll parameter $\epsilon_1$ remains very small 
throughout inflation.
So, we do not expect much change in the value of $\hnl$ from the case of 
de Sitter inflation and this expectation is confirmed by 
Fig.~\ref{fig:hnl-hi-ni}. 
For the sake of completeness we have included the plots of the scalar and 
tensor power spectra that arise in these models in Fig.~\ref{fig:ps}.
%%%%%%%%%%%%%%%%%%%%%%%%%%%%%%%%%%%%%%%%%%%%%%%%%%%%%%%%%%%%%%%%%%%%%%%%%%%%%%%
%%%%%%%%%%%%%%%%%%%%%%%%%%%%%%%%%%%%%%%%%%%%%%%%%%%%%%%%%%%%%%%%%%%%%%%%%%%%%%%%
%%%%%%%%%%%%%%%%%%%%%%%%%%%%%%%%%%%%%%%%%%%%%%%%%%%%%%%%%%%%%%%%%%%%%%%%%%%%%%%%
For the values of the parameters we have worked with, the hybrid inflation 
model seems to lead to a rather high tensor-to-scalar ratio, and hence it is
likely to be ruled out by the data.
In contrast, as we had mentioned, the aligned natural inflation model results
in a considerably suppressed tensor-to-scalar ratio and, therefore, it can 
be expected to be consistent with the data.

%%%%%%%%%%%%%%%%%%%%%%%%%%%%%%%%%%%%%%%%%%%%%%%%%%%%%%%%%%%%%%%%%%%%%%%%%%%%%%%%

\section{Consistency relation in the squeezed limit}\label{sec:sl}

\begin{figure}[!htb]
\begin{center}
\includegraphics[width=9.5cm]{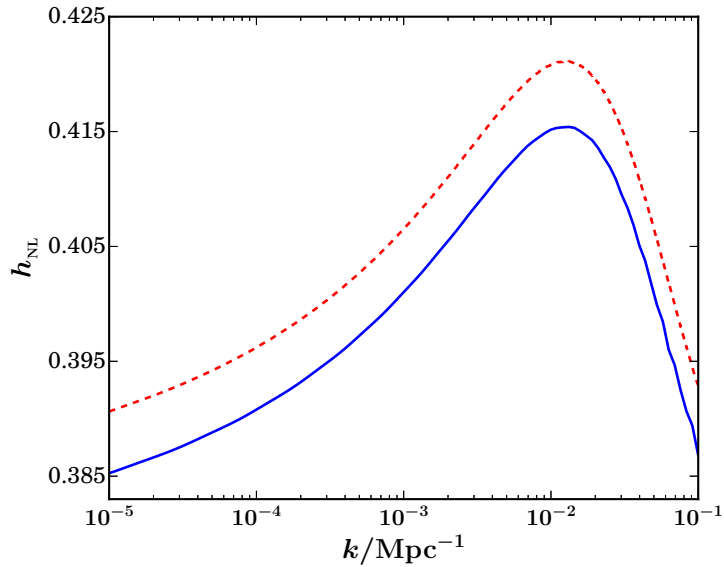}
\caption{\label{fig:hnl-sl}
The non-Gaussianity parameter $\hnl$ evaluated numerically in the squeezed 
limit (the solid line) and the quantity $(\nt-3)/8$ obtained from the
tensor power spectrum (the dashed line) have been plotted in the case of the 
double inflation model wherein $m_\chi=8\, m_\phi$.
The maximum difference between these quantities is about $1.2\%$ and 
the difference can be attributed to the level of numerical accuracy
that one has worked with.
This suggests that the consistency relation is valid even away from
slow roll.}
\end{center}
\end{figure}

It is well known that the amplitude of the tensor perturbations freeze
on super-Hubble scales.
Due to this reason, if one considers the long wavelength limit of one of
the wavenumbers (often referred to as the squeezed limit), it can be shown 
that the tensor bispectrum can be completely expressed in terms of the 
tensor power spectrum.
Specifically, if we choose $k_3\rightarrow 0$ so that $k_2\simeq k_3=k$, 
one finds that the non-Gaussianity parameter $\hnl$ can be expressed as 
follows~\cite{Sreenath:2014nka}:
\begin{equation}
\lim_{k_3\rightarrow 0} \hnl(\bm{k},-\bm{k},\vkc)=\f{\nt(k)-3}{8},
\label{eq:squeezed}
\end{equation}
where $\nt$ is the tensor spectral index defined as in Eq.~(\ref{eq:nt}).
Since we have been able to evaluate the non-Gaussianity parameter $\hnl$
(and the spectral index $\nt$) for an arbitrary triangular configuration 
of the wavenumbers, it is interesting to examine if the above consistency
is indeed satisfied in the models we have considered.
In Fig.~\ref{fig:hnl-sl}, we have plotted these two quantities for the
double inflation model with $m_\chi=8\,m_\phi$, which leads to the maximum 
possible deviation from slow roll.
%%%%%%%%%%%%%%%%%%%%%%%%%%%%%%%%%%%%%%%%%%%%%%%%%%%%%%%%%%%%%%%%%%%%%%%%%%%%%%%%
%%%%%%%%%%%%%%%%%%%%%%%%%%%%%%%%%%%%%%%%%%%%%%%%%%%%%%%%%%%%%%%%%%%%%%%%%%%%%%%%
We find that the maximum difference between these quantities evaluated
numerically is about $1.2\%$, which clearly supports the validity of
the consistency relation even in situations involving departures from
slow roll.
%%%%%%%%%%%%%%%%%%%%%%%%%%%%%%%%%%%%%%%%%%%%%%%%%%%%%%%%%%%%%%%%%%%%%%%%%%%%%%%%

\section{The contribution during preheating}\label{sec:cdp}

\begin{figure}[!htb]
\begin{center}
\includegraphics[width=9.5cm]{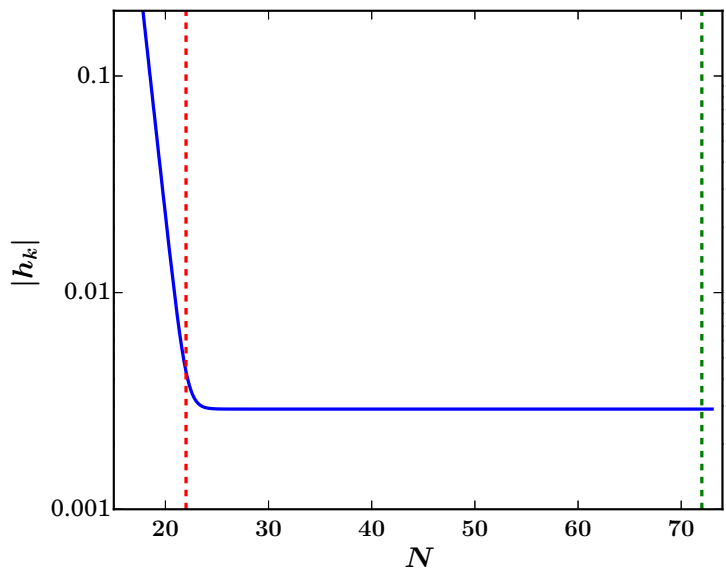}
\caption{\label{fig:hk}
Evolution of the absolute value of $h_k$ for $k=0.05\, {\rm Mpc}^{-1}$ 
during the epochs of inflation and reheating in the case of the double
inflation model with $m_\chi=8\,m_\phi$.
The vertical red and green lines indicate the time when the mode leaves
the Hubble radius during inflation and the end of inflation, respectively.}
\end{center}
\end{figure}

In models such as double inflation, the scalar field rolls down the potential 
and inflation is terminated when the field is close to the minimum of the 
potential. 
After inflation has ended, the scalar field oscillates about the minimum of 
the potential, a phase which is referred to as preheating.  
Note that all perturbations of cosmological interest are on super-Hubble scales 
during the domain of preheating.
Due to this reason, the oscillations in the scalar field are not expected
to affect the evolution of the amplitude of $h_k$, which remain constant
as in the super-Hubble domain during inflation.
We have evolved the tensor perturbations numerically through the epoch of 
preheating.
In Fig.~\ref{fig:hk}, we have plotted the evolution of the amplitude of
$h_k$ (for a specific mode) in the case of the double inflation model 
during the epochs of inflation and preheating.
Clearly, the figure corroborates the expectation that the amplitude of
$h_k$ is constant at suitably late times. 
%%%%%%%%%%%%%%%%%%%%%%%%%%%%%%%%%%%%%%%%%%%%%%%%%%%%%%%%%%%%%%%%%%%%%%%%%%%%%%%
%%%%%%%%%%%%%%%%%%%%%%%%%%%%%%%%%%%%%%%%%%%%%%%%%%%%%%%%%%%%%%%%%%%%%%%%%%%%%%%
Since the amplitude of the tensor modes is constant, the contribution to 
the non-Gaussianity parameter $\hnl$ due to this epoch is identically 
zero~\cite{Sreenath:2013xra}.  

%%%%%%%%%%%%%%%%%%%%%%%%%%%%%%%%%%%%%%%%%%%%%%%%%%%%%%%%%%%%%%%%%%%%%%%%%%%%%%%

\section{Discussion}\label{sec:d}

As we have stressed earlier, primordial non-Gaussianities are expected 
to provide crucial information to help us arrive at stronger constraints 
on the physics of the early universe.
Apart from the very recent efforts, there has been little work towards 
the numerical evaluation of non-Gaussianities in multi field models of 
inflation.
As a preliminary step, in this work, we have evaluated the tensor bispectrum
in two field models of inflation.
We have been able to compare the numerical results with the analytical
results available in the case of slow roll inflation.
This comparison suggests that the numerical procedure we have adopted
is quite accurate.
The two field models are interesting because of the curved trajectories 
that can be generated in the field space in a rather simple manner. 
One of our aims was to identify the effect of such a turn in the 
trajectory on the magnitude and the shape of $\hnl$. 
In double inflation, we have found that the change in the direction of the 
trajectory produces a bump in the first slow roll parameter, which increases 
the amplitude of $\hnl$ over a certain domain. 
We have also studied the behavior of $\hnl$ in the case of hybrid inflation 
and aligned natural inflation. 
Lastly, we have shown that the contribution to $\hnl$ due to the epoch of
preheating can be completely neglected, due to the constant amplitude of 
the tensor modes during this period.

\par

Evaluating the tensor bispectrum has proved to be simpler since the 
evolution of the tensor modes depend only on the behavior of the 
scale factor.
Moreover, the fact the tensor modes freeze on super-Hubble scales makes
the computation easier.
We are presently extending our code to evaluate the other three-point
functions in two field models.
In the case of the three-point functions involving scalars, the presence 
of iso-curvature perturbations provides a challenge, as they can lead 
to non-trivial evolution of the curvature perturbation on super-Hubble 
scales.
We are currently working on this issue.

%%%%%%%%%%%%%%%%%%%%%%%%%%%%%%%%%%%%%%%%%%%%%%%%%%%%%%%%%%%%%%%%%%%%%%%%%%%%%%%

\section*{Acknowledgements}

We wish to thank Debika Chowdhury for comments on the manuscript.
LS wishes to thank the Indian Institute of Technology Madras, Chennai, India, 
for support through the New Faculty Seed Grant.

%%%%%%%%%%%%%%%%%%%%%%%%%%%%%%%%%%%%%%%%%%%%%%%%%%%%%%%%%%%%%%%%%%%%%%%%%%%%%%%
\bibliographystyle{JHEP}
\bibliography{tensor-bispectrum-june-2017}
%%%%%%%%%%%%%%%%%%%%%%%%%%%%%%%%%%%%%%%%%%%%%%%%%%%%%%%%%%%%%%%%%%%%%%%%%%%%%%%

\end{document}